\documentclass[table]{ccjnl}
\usepackage{lipsum,amsmath}
\usepackage{cuted}
\usepackage{bm}
\usepackage{epstopdf}
\graphicspath{{figures/}}

\title{New Delay Doppler Communication Paradigm in 6G era: A Survey of Orthogonal Time Frequency Space (OTFS)}
\author{Weijie Yuan\inst{1}, Shuangyang Li\inst{2}, Zhiqiang Wei\inst{3}, Yuanhao Cui\inst{4}, Jiamo Jiang\inst{5}, Haijun Zhang\inst{6}, and Pingzhi Fan\inst{7}}
\receiveddate{Aug. 10 2022}
\reviseddate{Sep. 23 2022}
\Editor{}

\address[1]{Department of Electronic and Electrical Engineering, Southern University of Science and Technology, Shenzhen 518055, China}
\address[2]{Department of Electrical Engineering and Computer Science, Technical University of Berlin, Berlin 10623, Germany}
\address[3]{School of Mathematics and Statistics, Xi’an Jiaotong University, Xi’an 710049, China}
\address[4]{Department of Communication Engineering, Beijing University of Posts and Telecommunications, Beijing 100876, China}
\address[5]{China Academy of Information and Communications Technology, Beijing 100191, China}
\address[6]{School of Computer and Communication Engineering, University of Science and Technology Beijing, Beijing 100083, China}
\address[7]{School of Information Science and Technology, Southwest Jiaotong University, Chengdu 610031, China}

\begin{document}

\maketitle

\begin{abstract}
In the 6G era, space-air-Ground integrated networks (SAGIN) are anticipated to deliver global coverage, necessitating support for a diverse array of emerging applications in high-mobility, hostile environments. Under such conditions, conventional orthogonal frequency division multiplexing (OFDM) modulation, widely employed in cellular and Wi-Fi communication systems, experiences performance degradation due to significant Doppler shifts. To overcome this obstacle, a novel two-dimensional (2D) modulation approach, namely orthogonal time frequency space (OTFS), has emerged as a key enabler for future high-mobility use cases. Distinctively, OTFS modulates information within the delay-Doppler (DD) domain, as opposed to the time-frequency (TF) domain utilized by OFDM. This offers advantages such as Doppler and delay resilience, reduced signaling latency, a lower peak-to-average ratio (PAPR), and a reduced-complexity implementation. Recent studies further indicate that the direct interplay between information and the physical world in the DD domain positions OTFS as a promising waveform for achieving integrated sensing and communications (ISAC). In this article, we present an in-depth review of OTFS technology in the context of the 6G era, encompassing fundamentals, recent advancements, and future directions. Our objective is to provide a valuable resource for researchers engaged in the field of OTFS.
\end{abstract}
\section{Introduction}
\label{Introduction}
\begin{figure*}
 \centering
 \includegraphics[width=.85\textwidth]{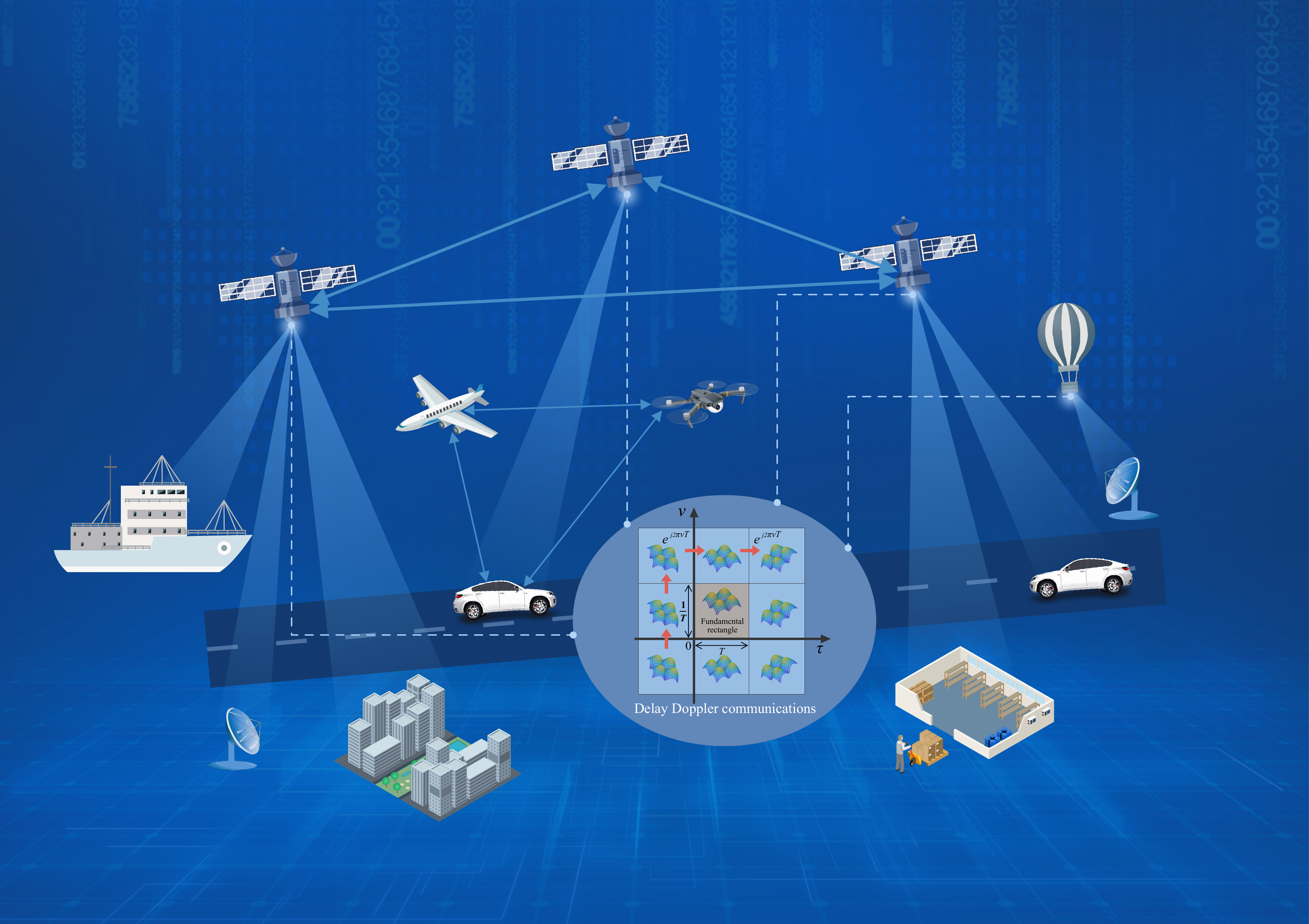}
 \caption{Space-air-ground integrated networks (SAGIN) relying on delay Doppler Communications}
	\label{sagin}
\end{figure*}

After years of research and development, the fifth-generation (5G) wireless systems have been standardized and commercialized recently globally\cite{andrews2014will}. Nowadays, both academia and industry have shifted their focus toward the development of sixth-generation (6G) technologies \cite{wang2022vision}. Hindered by the coverage and capacity limitations of traditional terrestrial wireless communications, the 5G system is unable to support high data rates and reliability universally, which is one of 6G wireless systems' primary objectives. The innovative network architecture, namely Space-Air-Ground Integrated Network (SAGIN), has been identified as a crucial enabler for 6G, facilitating seamless connectivity and high data rate transmission \cite{liu2018space}. The realization of SAGIN will involve a wide range of emerging applications, e.g., mobile communications on board Aircraft (MCA), low-earth-orbit satellites (LEOS), self-driving autonomous cars, in-vehicle infotainment, and unmanned aerial vehicles (UAV) \cite{su2019broadband,zhou2020evolutionary,gupta2015survey}, as shown in Fig. \ref{sagin}. In the above use cases, a critical challenge is how to provide reliable communication in high-mobility environments. For example, the relative speeds for vehicle-to-vehicle (V2V) communications will be up to $300$ km/h. In high-speed railways (HSR) mobile service, the communication devices will have speeds of up to 500 km/h. For MCA and LEO satellite communications, the moving speeds of the user equipments (UEs) will be even higher. As discussed in the 6G vision paper by You. et.al, high-mobility is still a bottleneck of the traditional wireless systems because of severe Doppler spread effect caused by the relative motion between transceivers. Moreover, ultra-high data rate requirements prompt mobile providers to exploit higher frequency bands, such as millimeter-wave (mmWave) bands, where a vast spectrum is available \cite{li2018mmwave,wang2018millimeter}. Within the mmWave frequency band, even minor UE speeds can result in significant Doppler shifts.

The widely-used orthogonal frequency division multiplexing (OFDM) technology has proven to be an efficient solution for addressing inter-symbol interference (ISI) caused by time dispersion, a result of the multipath effect. By adopting the cyclic prefix (CP), OFDM can overcome ISI in 4G and 5G systems \cite{li2006orthogonal}. However, it will fail to support efficient and reliable communications in doubly selective channels, where additional frequency dispersion is caused by Doppler shifts. This can be explained by the fact that high Doppler shift will result in a very short channel coherence time, which destroys the orthogonality between sub-carriers of OFDM and leads to inter-carrier interference (ICI). Against this background, designing new modulation techniques and waveforms, as well as the corresponding transceiver architectures, to achieve high-speed and ultra-reliable communications in challenging environments becomes urgent and paramount for future mobile systems \cite{zheng2017comparison}.

The orthogonal time-frequency space (OTFS) technique, which modulates data in the delay-Doppler (DD) domain rather than the conventional time-frequency (TF) domain, was initially proposed for high-mobility wireless applications and has been recognized globally as a ground-breaking technology and an enabler for future wireless communications \cite{hadani2017orthogonal,wei2021orthogonal}. Although the term “OTFS” was first introduced in 2017, preliminary research on channel characteristics in the DD domain can be traced back to 1960s \cite{bello1963characterization}. Relying on the DD domain signal representation, OTFS provides an innovative framework for investigating the interaction between information symbols and wireless channels, which allows the benefits of strong Doppler-resilience and delay-resilience against highly dynamic and complex environments \cite{monk2016otfs}. More importantly, the unitary transformation from the DD domain to the TF domain in OTFS spreads each information symbol to the entire TF domain. Thus, the DD domain symbol is capable of experiencing the full TF channel, providing full time-frequency diversity with appropriate receiver design, which is crucial for reliable communications \cite{wei2022otfstutorial,yuan2023orthogonal}. Another benefit of DD communication waveform is the quasi-periodicity, which is also illustrated in Fig. \ref{sagin}. In fact, the DD waveform is periodic up to a multiplicative phase, which indicates that OTFS-based DD communications can adaptively support the transceivers with different speeds. Additionally, the DD domain channel exhibits desired properties such as sparsity and stability, which can be leveraged for accurate channel estimation with low training overhead and low-complexity signal detection. Furthermore, OTFS offers more appealing advantages over OFDM, including a lower peak-to-average power ratio (PAPR), reduced signaling overhead due to the reduced cyclic prefix frame structure, and enhanced robustness against synchronization errors. These advantages make OTFS an excellent enabler for hostile wireless applications in the 6G era \cite{li2021tutorial}.

In addition to communications, it is well agreed that the sensing capability will play a more significant role than ever before in 6G era. Consequently, there is a growing trend towards integrating both functionalities by jointly designing waveforms and transceiver architectures, known as integrated sensing and communication (ISAC). As for radio sensing, it relies on range measurements for positioning the target and the estimation of the target speed. In theory, the range can be readily converted from the signal propagation delay while the speed can be obtained by giving the observed Doppler shift. Obviously, sensing inherently involves the acquisition of delay and Doppler parameters, which perfectly aligns with OTFS's utilization of the DD domain for communication purposes. As discussed in \cite{monk2016otfs}, the DD communication channel also reveals the underlying physical propagation environments. Therefore, although the original purpose for developing OTFS is to support high-mobility applications, the characterization of channel in the DD domain pave the way to approach the physical world. OTFS has been widely acknowledged as a promising waveform for realizing ISAC by providing a novel framework to unify both functionalities in the same DD domain \cite{yuan2022otfstutorial}.

Considering quite a few benefits provided by OTFS, the interests from both the industry and academia have been tremendously increased since 2017. In 2021, a special issue on OTFS was published in ZTE Communications \cite{yuan2021special}, which has attracted several contributions. As for the overview paper, there has been one published recently in IEEE Internet of Things (IoT) Journal, which, however, focused on the application of OTFS in IoT networks \cite{xiao2021overview}. In this article, we aim to present a comprehensive survey of OTFS technology in 6G era, including the historical views, fundamentals, recent advances, and future works. We hope this article will help all researchers, especially for those new in this area to have a panorama of OTFS.

\section{OTFS Principles}
\label{OFDM}
\subsection{Wireless Channels in Delay Doppler (DD) Domain}
Before discussing OTFS, we first consider the wireless channels, which play a fundamental role in wireless communications.
The time and frequency domain is well acknowledged for channel representation thanks to the success of OFDM in the currently deployed communication systems. Based on the Fourier transformation from time domain (TD) to frequency domain (FD), the frequency-selective fading channel can be converted to independent FD sub-channels, which motivates the data symbol multiplexing in FD and the implementation of OFDM. In fact, this can be explained by that the eigenvalues of the frequency-selective fading channel are the corresponding channel frequency responses. However, the existence of eigenvalue decomposition of communication channel in FD only holds for static or quasi-static channel conditions. As for the time-selective fading channel or doubly-selective channel with Doppler shift, OFDM may fail to work since the FD sub-channels are no longer independent. Naturally, we will ask whether there exist another domain where the eigenvalue decomposition of doubly-selective channel exist. Unfortunately, the answer is no.

For the purpose of designing a new signaling waveform, how to combat the channel fading is essential. By revisiting the channel fading, we summarize that the time and frequency-selective fading are the consequences of wireless transmission with respect to the delay and Doppler shifts. Therefore, it is possible to adopt the delay and Doppler parameters for channel representation, instead of the commonly used time-frequency (TF) parameters. In Bello's pioneering paper \cite{bello1963characterization}, he provided the mathematical fundamentals of randomly linear time-variant channels. Both the wide-sense stationary (WSS) channel and the uncorrelated scattering (US) channel were introduced, which are shown to be time-frequency duals. Then, specific focus is given to WSSUS channels, whose scatter function can be fully characterized by using time and frequency variables or delay and Doppler variables. In fact, the physical attributes of the channel remain roughly unchanged during the signal transmission, resulting in a time-invariant DD domain channel, compared to its time-varying TF domain counterpart. In addition to time invariance, the DD channel representation is inherently sparse, since it is determined by the number of scatterers during the signaling transmission. Relying on this property, the work of \cite{li2007estimation} considers the DD domain-based channel in oceanic acoustic environments. It utilizes the DD spread function for time-varying channel representation and develops a channel estimation algorithm. The proposed approach can capture both the acoustic channel structure and its dynamics without explicit dynamic channel modeling. Besides the theoretical analysis, Molisch et.al have conducted experiments in 60GHz vehicle-to-infrastructure communication channels in a street crossing scenario in an urban environment and showed that the DD channels are indeed sparse under the considered communication scenarios \cite{groll2019sparsity}.

A key feature of WSSUS channels is that the delay and Doppler responses remain unchanged~\cite{Matz2005nonWSSUS}. The physical interpretation for this feature is that the delay and Doppler relate to the physical attributes of the channel scatters, e.g., relative distance and velocity, which will not change if the channel geometry remains static. More generally, non-WSSUS channels consider the signal transmission over a channel whose geometry could change, leading to the variation of delay and Doppler. It turns out that there is a stationarity region for general non-WSSUS channels, within which the non-WSSUS channel can be roughly viewed as a WSSUS channel. As such, the non-WSSUS channel has roughly unchanged delay and Doppler responses within each stationarity region~\cite{hlawatsch2011wireless}, which is similar to the coherence region for TF domain channels. More importantly, the size of the stationarity region is inversely proportional to the maximum delay and Doppler correlation lag, i.e., the maximum delay and Doppler lags corresponding to which the channel correlation function is non-zero, which is much larger than the typical coherence region in the TF domain~\cite{Matz2005nonWSSUS}. In fact, many numerical simulations for DD domain communication systems consider non-WSSUS channels, where the delay and Doppler responses are generally assumed to be constant within in one or several consecutive frames (corresponding to one stationarity region), while the responses change independently across different stationarity regions.

\begin{figure}
 \centering
 \includegraphics[width=.5\textwidth]{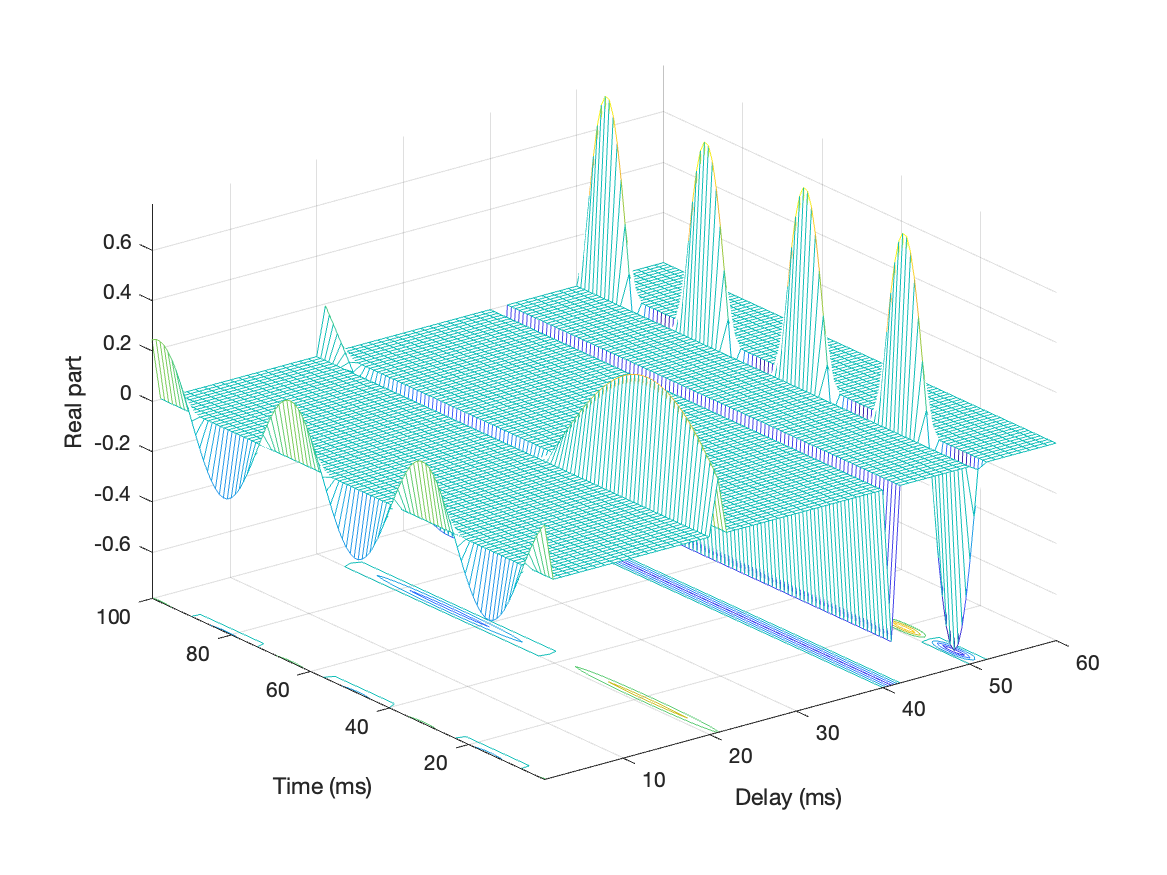}
 \caption{Time-delay domain effective channel}
	\label{td}
\end{figure}
\begin{figure}
 \centering
 \includegraphics[width=.5\textwidth]{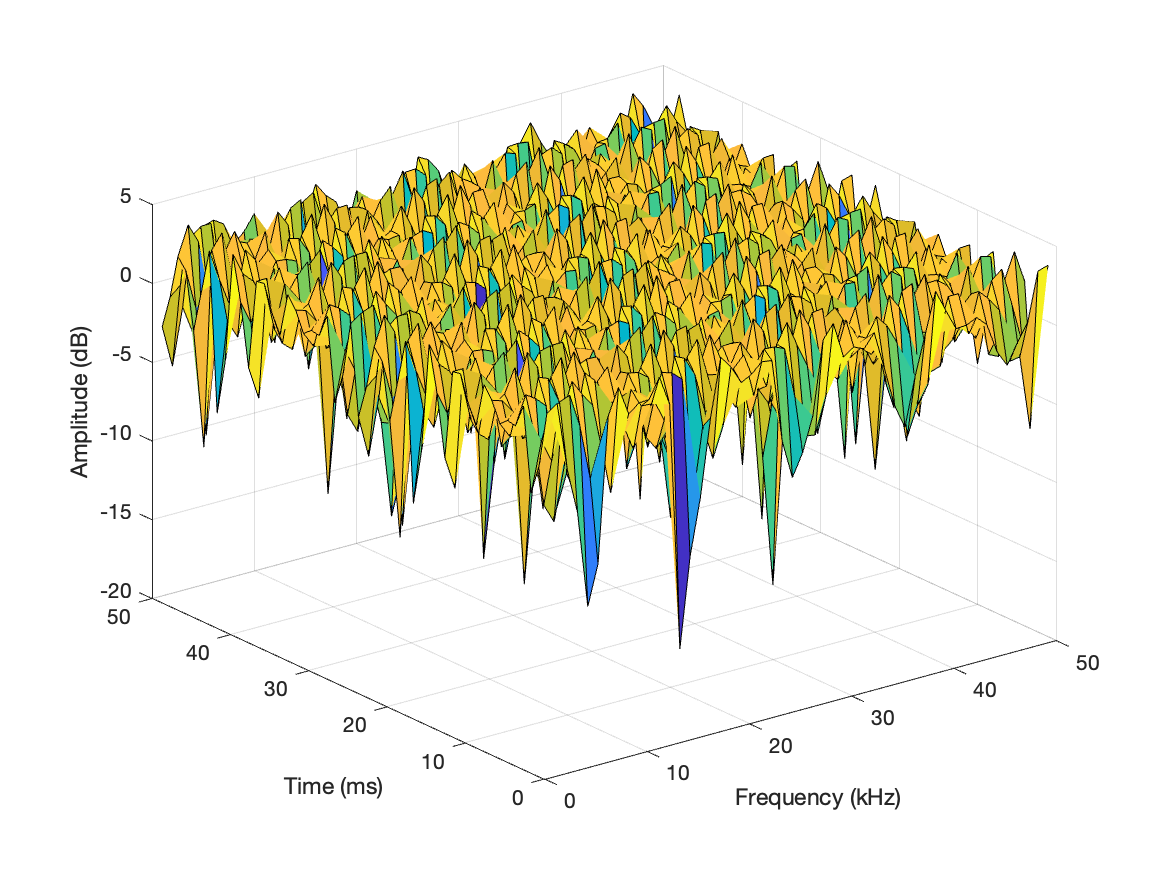}
 \caption{Time frequency domain effective channel}
	\label{tf}
\end{figure}
\begin{figure}
 \centering
 \includegraphics[width=.5\textwidth]{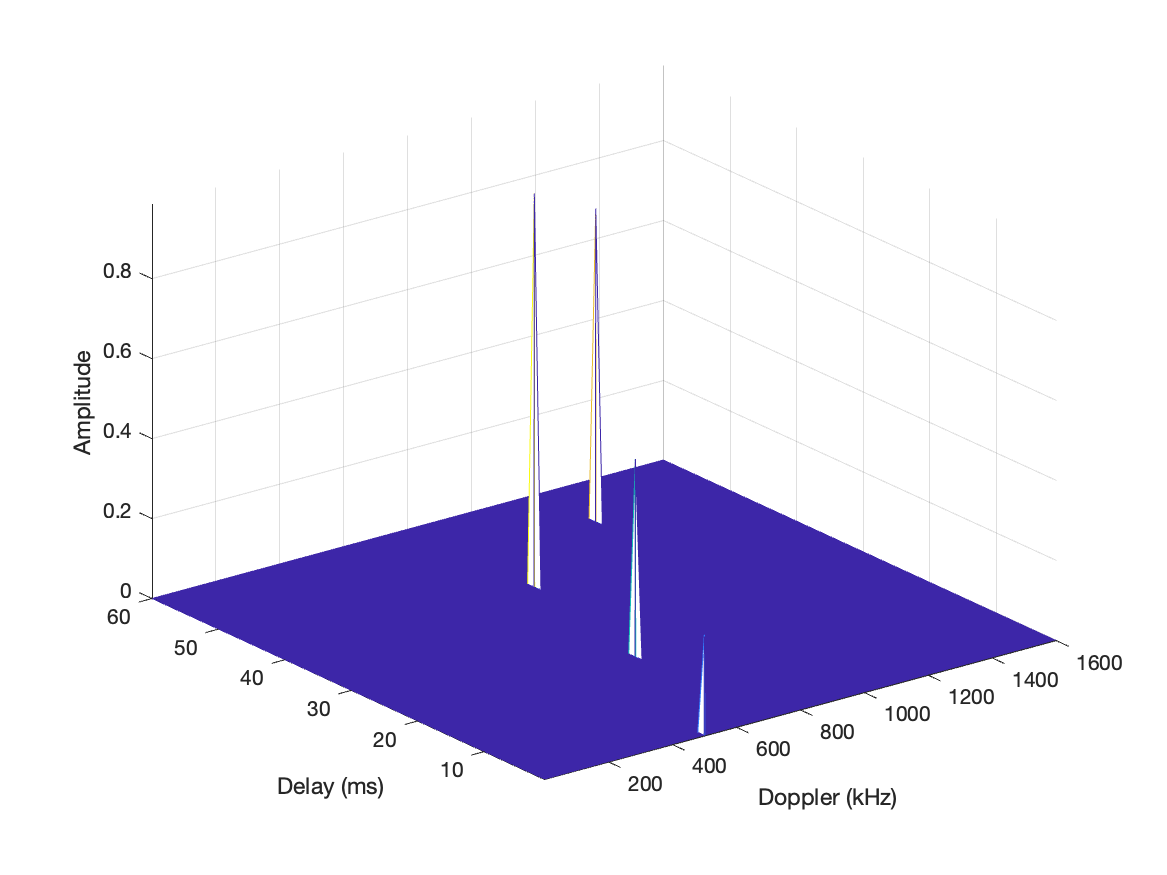}
 \caption{Delay Doppler domain effective channel}
	\label{dd}
\end{figure}

For ease of exposition, in Fig. \ref{td} - \ref{dd} we illustrate the channel representations within the stationarity region in the time-delay domain, TF domain, and DD domain, respectively. Observed from Fig. \ref{td} - \ref{dd}, several attractive benefits are enjoyed by DD domain channel representation, which are summarized as follows
\begin{itemize}
	\item \textbf{Stability}: The DD domain channel can be seen as a `snapshot' of the real wireless environment. Therefore, only burst changes of propagation path lengths and moving speeds will cause variations of the DD parameters, leading to much slower channel fluctuations.
	\item \textbf{Sparsity}: The common wireless propagation environment has only a limited number of moving scatterers/reflectors, exhibiting a sparse channel response in the DD domain.
	\item \textbf{Separability}: Conventionally, the paths with the same delay are not separable as the signal arrives at the receiver simultaneously. Nevertheless, with the aid of Doppler information, we can separate different paths and thus enjoy full TF diversity.
	\item \textbf{Compactness}: All channel responses only appear in a DD domain region bounded by the maximum delay and Doppler, which are associated with the maximum range distance and moving speed in the wireless environments, regardless of OTFS frame size.
\end{itemize}
The DD channel representation has successfully connected the physical and the `digital' worlds, which motives the development of OTFS.

\subsection{OTFS Concepts}
The term \emph{Orthogonal Time Frequency Space (OTFS)} was first introduced in 2017 in a conference paper by Hadani et.al \cite{hadani2017orthogonal}. This paper commenced from the DD domain signal representation point of view and compares the performance of OTFS to the classic OFDM modulation. The merits of OTFS, including its conceptual connection with Radar, new coupling relationship between information and channels, robustness against interference, and linear scaling of spectral efficiency with respect to the number of antennas are revealed. Later in the white paper by Cohere Technologies \cite{hadani2018otfs}, the authors reveal its connections to conventional modulation waveforms, such as time division multiple access (TDMA), OFDM, and code division multiple access (CDMA).

\begin{figure*}
 \centering
 \includegraphics[width=.8\textwidth]{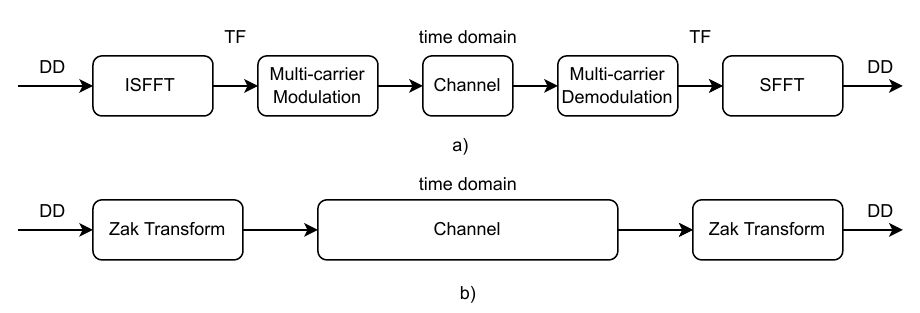}
 \caption{Block diagram of OTFS implementation: a) OFDM-based two-step conversion; b) Zak transform-based one-step conversion.}
	\label{fig1}
\end{figure*}

The main idea of OTFS is to multiplex the transmitted data symbols in the DD domain. Fig. \ref{fig1} depicts the original implementation block diagram of OTFS. Firstly, the information bits are mapped to data symbols and each symbol will occupy a DD grid with delay index $l$ and Doppler index $k$. The size of a DD grid depends on the available bandwidth and signal time duration. Secondly, the inverse symplectic finite Fourier transform (ISFFT) is performed to convert the DD signal to the TF domain. Then, conventional multi-carrier modulators, e.g., OFDM, transforms the signal into the time domain. After traveling through the wireless channel, the time domain received signal is converted to the TF domain using a multi-carrier demodulator and finally is transformed to the DD domain via SFFT. In \cite{farhang2017low}, the authors derived a discrete-time formulation of an OTFS system based on OFDM transceiver architecture. The authors further studied simplified modulator and demodulator structures to provide deeper insights into OTFS systems. Instead of the commonly used Fourier transform, the work \cite{mallaiah2022novel} further designed an OTFS system design using the discrete fractional Fourier Transform (FrFT)-based OFDM system. The FrFT technique helps to achieve a better performance and a lower PAPR compared to the conventional OTFS system at the same order of complexity. All the above OTFS system implementations adopted a two-step domain transformation, i.e. from DD to TF and to the time domain at the transmitter side and vice versa for the receiver, which is mainly due to the compatibility requirement with OFDM technology.

Considering the intrinsic connection between time domain and DD domain signals via the sophisticated Zak transformation (ZT), the OTFS signal can be directly converted to time domain while bypassing the TF processing. In \cite{lampel2021orthogonal}, the authors introduced the discrete ZT (DZT) approach to implement OTFS. After reviewing the properties of DZT, the connections between OTFS and conventional OFDM were investigated. In particular, the overlay between the pulse shaping OFDM and OTFS for both rectangular and non-rectangular pulses are investigated. Then, a simple implementation of OTFS is proposed by leveraging DZT, and a concise input-output relationship in the presence of practical wireless channels is also derived. Another work by Saif Mohammed provided an interesting interpolation of OTFS based on ZT \cite{mohammed2021derivation}. This paper rigorously derived an orthonormal basis of approximately time and bandwidth-limited signals which are also localized in the DD domain. Furthermore, Saif showed that irrespective of the amount of Doppler shift, the received DD domain basis signals are localized in the DD domain w.r.t the DD resolution. Finally, this paper verifies that the degree of localization of the DD domain basis signals is inversely related to bandwidth and time duration of the transmit signal. Compared to the two-step DD-TF-TD domain conversion, using Zak transformation was shown to have a lower implementation complexity \cite{mohammed2021time,wei2022otfstutorial}.

\subsection{OTFS Fundamentals and Performance Limits}
As a new modulation waveform in the DD domain, several pieces of research focus on the fundamental performance of OTFS from different points of view. The work of \cite{biglieri2019error} examined the performance of OTFS, especially the amount of achievable diversity over fading channels. As one of the benefits of OTFS, the separability of channel paths in the Doppler domain provides the potentials of achieving full TF diversity. Then, the work \cite{surabhi2019diversity} showed that the asymptotic diversity order of OTFS modulation is one. However, the potential for a higher order diversity is witnessed before the diversity one regime takes over in the finite signal-to-noise ratio (SNR) regime. Furthermore, a simple phase rotation scheme for OTFS was proposed using transcendental numbers, based on which OTFS transmission is guaranteed to achieve the full diversity \cite{surabhi2019diversity}. The paper by Raviteja et.al proved that the full effective diversity, i.e., the slope of the majority of pairwise error probability (PEP) curves rather than the minimum one, can be achieved by using OTFS modulation with practical shaping pulses \cite{raviteja2019effective}. The authors drew a conclusion that OTFS practically achieves full effective diversity with sufficiently large signal constellations. Previous works mostly assumed ideal pulse shaping for diversity analysis. The authors of \cite{wang2022transmit} studied the practicability of previous diversity schemes in the rectangular pulse-based OTFS systems and developed a diversity achieving scheme. To do so, the information symbols were divided into two half frames along the Doppler domain, which share identical equivalent channels.

The high peak-to-average ratio (PAPR) issue is vital for communication systems with power constraints or nonlinear High-Power-Amplifier (HPA). For OFDM modulation, the PAPR is proportional to the number of sub-carriers, which may become very high due to the large number of sub-carriers \cite{wulich2005definition}. However, as discussed in \cite{surabhi2019peak}, the PAPR for OTFS is proportional to the number of Doppler bins/time slots, which is much smaller than the OFDM counterpart. Nevertheless, there is still a demand to further decrease the PAPR in OTFS system. In \cite{wei2021charactering}, the asymptotic expression of the PAPR distribution was derived, based on the assumption of the infinity-approaching number of data symbols in the DD domain. It was also shown that the derived analytical PAPR expression can be also extended to the case of a bandlimited pulse, by employing the properties of the hi-squared process. The authors of \cite{hossain2020dft} analyzed the PAPR of the OTFS signaling and proposed a new concept of DFT-spread OTFS, which can efficiently reduce the PAPR. The main idea of the DFT-spread OTFS is to average the energy of a single subcarrier. Simulation results have shown significant BER performance gain in HPA nonlinear conditions. Later, the authors from Xidian Uni. proposed an iterative clipping and filtering (ICF) framework for PAPR reduction. Based on the proposed framework, it is capable of striking a good trade-off between PAPR and BER performance \cite{gao2020peak}. An autoencoder (AE) architecture-based PAPR reduction method was developed \cite{liu2021autoencoder}, where the encoder part is trained for reducing the PAPR while the decoder part is used for signal reconstruction. To achieve precise control of peak-to-average power ratio (PAPR) and BER performance, a recent work \cite{wang2022performance} proposed a nonlinear corrective active constellation expansion method in OTFS waveform. Different from the linear clipping and filtering process, nonlinear compressed clipping noise is considered with a flexible curvature control coefficient, providing the capability of striking a trade-off between PAPR and BER performances.

As for generalized performance analysis, the authors of \cite{raviteja2019otfs} showed the performance gain of OTFS in static multi-path channels. From the capacity perspective, the authors of \cite{gaudio2021otfs} studied the pragmatic capacity of both OTFS and OFDM in the presence of sparse channels. The pragmatic capacity is defined as the achievable rate of the channel induced by the signal constellation and the detector soft-output. In particular, this paper demonstrated that OTFS outperforms OFDM over static channels under practical channel estimation and detection approaches while OTFS enjoys a more robust performance than OFDM over high-mobility channels. The effect of power allocation (PA) algorithms on OTFS BER performance was analyzed in \cite{wang2023ber}. Two different PA algorithms in OTFS systems with zero forcing (ZF) and minimum mean square error (MMSE) equalization, and a suboptimal MBER power allocation method were developed to achieve better BER performance. The authors also raised a combined subsymbol allocation (SA) and PA strategy in the case of MMSE equalization and then investigate the BER performance with the proposed joint SA and PA strategy in delay-Doppler channels. To provide the guidelines of code design in OTFS system, \cite{li2021performance} investigated the error performance of coded OTFS systems. Although the diversity gain of OTFS systems improves with the number of resolvable paths in the DD domain, the coding gain declines. Rule-of-thumb code design criterion was provided in this paper, which is increasing the Euclidean distance between codeword pairs. The performance of OTFS system with antenna selection at the transmitter and the receiver was analyzed in \cite{bhat2021performance} and \cite{bhat2021performancetwc}, respectively, which showed that the phase rotation is a necessity to extract full diversity. Moreover, it is able to predict the diversity orders through the rank of the different matrices. The in-phase and quadrature (IQ) imbalance will also impact the performance of communications systems. To this end, \cite{tusha2021performance} investigated the performance of OTFS transmission in the presence of IQ impairment. Mirror-Doppler interference (MDI), which only appears in the Doppler domain was derived to illustrate the physical effect of the IQ mismatched transmitter. An interesting conclusion is that IQ imbalance is critical to the diversity gain of OTFS system but does not lead to a BER error floor. In addition to IQ imbalance, the work \cite{neelam2022analysis} further considered impairments such as carrier frequency offset (CFO) and direct current (DC) offset. The DD input-output model under these impairments was first derived, based on which the impairment parameters were estimated in the DD domain while were compensated in the time domain.

\section{Transceiver Design for OTFS System}
\label{PROPOSED}
Since OTFS was first introduced, the design of the transmitter and receiver has become a popular research topic. In this section, we will overview the recent advances on the topic of transceiver design for OTFS \cite{li2022otfstutorial}.

\subsection{Transmitter Precoding Design}
The choice of shaping pulse affects the performance of OTFS system. The ideal pulse, which was assumed in most works, can satisfy the bi-orthogonality property but can not be realized in practice. The paper \cite{raviteja2018practical} analyzed the input–output relation of OTFS system for arbitrary pulse-shaping waveforms using a block-circulant matrix decomposition. The out-of-band radiation and BER performance with different waveforms were also compared. The work of \cite{lin2022multicarrier} developed a multi-carrier modulation scheme based on practical pulse shaping and named the waveform as orthogonal delay-Doppler division multiplexing (ODDM), which also relies on the DD grids for data transmission.

Alternative to directly designing the shaping pulse in the time domain, one can design the window function in the TF domain. The work by Wei et. al investigated the impacts of window functions for OTFS modulation \cite{wei2021transmitter}. The TX window can be interpreted as a power allocation in the TF domain while employing an RX window causes a colored noise. When channel state information (CSI) is available at both the TX and RX, an optimal TX window can be designed to minimize the detection performance. When CSI is not available at the TX, a proper window design, such as Dolph-Chebyshev (DC) window, in the TF domain can enhance the channel sparsity when the Doppler/delay resolution is insufficient \cite{wei2021performance}. A similar work considered frequency domain precoding was devised in \cite{zhang2021adaptive}. In this paper, the authors proposed an adaptive transmission scheme with frequency domain precoding that is composed of the eigenvalues of the channel matrix to improve the MMSE detection performance. As a further step, the lower and upper bounds for the diversity performance of the adaptive transmission scheme were derived. Considering the precoding in the DD domain, the work of \cite{sun2022closed} developed a precoder by optimally allocating the power to each DD domain grid. Based on the closed-form BER expression, a lower BER is achieved with designed precoders.

For multi-user and multi-antenna case, research on transmit precoding design for OTFS system has just commenced. In \cite{pandey2021low}, a multi-user precoder at the base station (BS) and a corresponding low complexity detector at the user side were proposed. Relying on the proposed precoder and detector, it is capable of separating the demodulation of different OTFS symbols at the UT. The complexity of the proposed precoder only increases with the number of BS antennas and the number of users. Two low-complexity precoding algorithms for MU MIMO-OTFS were developed in \cite{cao2021low}, where the high-dimensional matrix inversion is done by adopting the Woodbury matrix identity, a.k.a, matrix inversion lemma. With the aid of the proposed precoding algorithm, the performance of the considered system was robust to different user speeds. A recent work on air-to-ground (A2G) communications developed a low-complexity hybrid precoding algorithm with rectangular waveforms for mmWave-OTFS system \cite{yan2022low}. The DD domain signal transmission process was first established, followed by the derivation of the DD domain channel matrix. Finally, the low-complexity hybrid precoding algorithm was proposed by adopting a matrix approximation-based strategy.

\subsection{Delay-Doppler Channel Estimation}
\label{Design}
Accurate channel acquisition and estimation are critical for reliably detecting the transmitted data symbols. By leveraging the fact that the received DD symbols can be represented by two-dimensional (2D) circulant convolution of the transmitted symbols and effective DD channel, the authors of \cite{murali2018otfs} proposed a pseudo-random noise (PN) pilot-based channel estimation scheme in 2018. To improve the channel estimation accuracy, the paper \cite{raviteja2019embedded} proposed a pilot-aided DD domain channel estimation scheme, where a single pilot impulse with guard around it is inserted in the DD domain. A threshold-based method was employed to estimate the channel with both integer and fractional Doppler shifts. The extension of the proposed scheme to MIMO and multi-user cases was also discussed. Liu et al. in \cite{liu2023low} investigated the problem that channel estimation schemes may result in poor PAPR performance of OTFS system or low spectrum efficiency. Therefore, the authors proposed a low PAPR channel estimation scheme with high spectrum efficiency. Specifically, a multiple scattered pilot pattern was designed, where multiple low-power pilot symbols were superimposed with data symbols in the DD domain. It is shown that the proposed scheme can significantly reduce the PAPR while keeping the spectrum efficiency.

As for the estimation problem, Cram\'er-Rao lower bound (CRLB) serves as a benchmark \cite{han2015performance,xiong2017cooperative}. In \cite{pan2021cramer}, the authors formulated an exact expression of the discrete-time system model for non-ideal pulse-shaping OTFS system and derived the CRLB of OTFS channel estimation in different cases. Considering the leakage issue which reduces the measurement precision, \cite{pfadler2022leakage} proposed a new channel estimation scheme that exploits smoothness regularization in the TF domain to suppress the leakage in the DD domain.

Since the DD channel reflects the actual wireless propagation geometry, some sparse estimation techniques based on compressive sensing have been adopted for capturing the channel sparsity. In \cite{shen2019channel}, a structured orthogonal matching pursuit algorithm-based channel estimation scheme for massive MIMO-OTFS systems was proposed by exploiting the normal sparsity along the delay dimension, block sparsity along the Doppler dimension, and burst sparsity along the angle dimension. Numerical results have shown a superior channel estimation performance achieved by the proposed scheme. In \cite{liu2020uplink}, the authors adopted the expectation maximization-based variational Bayesian (EM-VB) framework for recovering the uplink DD channel. Relying on the slow time variance of the DD channel, the authors further exploited the delay and Doppler reciprocity between the uplink and the downlink and reconstruct the associated downlink channel parameters. The work of \cite{liu2021message} formulated the DD channel estimation as a structured signal recovery problem. With a factor graph representation of the problem, a message passing algorithm was developed to estimate the channel gains and fractional Doppler shifts jointly. The effectiveness of the proposed algorithm is verified by comparing the estimation error with the derived CRLB. This paper showed that the proposed algorithm is able to work with multiple pilot symbols to achieve significant PAPR reduction. Based on \cite{liu2021message}, Li et.al adopted the hidden Markov model (HDD) as the prior of the hyper-parameter and developed a message passing-based channel estimation scheme \cite{li2022iccc}. The authors of \cite{zhao2022block} proposed block sparse Bayesian learning (SBL) with block reorganization (BSBL-BR) method to recover the channel information. Compared to the classic BSBL method, the proposed algorithm iteratively updated the size of non-sparse blocks, leading to better channel estimation accuracy. Aiming for addressing the challenges caused by fractional inter-Doppler interference (IDI), the authors of \cite{liao2023joint} derived the original channel response (OCR) from the time domain channel impulse response (CIR) to reduce the dimension of OCR by using the basis expansion model (BEM) and the relationship between the time and DD domain channel. SBL was used to estimate the basis coefficients in the BEM without any \emph{a priori} information of the channel. To accurately determine the fractional Doppler shift/delay caused by insufficient signal duration/bandwidth, \cite{wei2022off} proposed an off-grid channel estimation scheme for OTFS systems, which estimates the original DD domain channel response rather than the effective DD domain channel response as commonly adopted in the literature. Two problems, i.e., one-dimensional (1D) and two-dimensional (2D) off-grid sparse signal recovery problems were formulated and solved by SBL-based algorithms. The on-grid delay and Doppler shifts were jointly determined by the entry indices with significant values in the recovered sparse vector. Then, the corresponding off-grid delay and Doppler shifts were modeled as hyper-parameters in the proposed SBL framework, which can be estimated via the expectation-maximization method. To tackle the non-Gaussian noise in some complex environments, the authors of \cite{zhang2022iccc} developed a deep residual learning framework for OTFS channel estimation, which is proven to efficiently estimate the DD channel with arbitrary measurement noise. 

Conventionally, the guard space inserted in the DD domain which avoids the interference between data symbols and pilot is utilized to guarantee the channel estimation accuracy. However, the size of guard space depends on the maximum delay and Doppler indices, which undoubtedly deteriorate the spectral efficiency. To this end, the pilot design has also attracted many interests. A new pilot pattern without guard space was proposed in \cite{zhao2020sparse}, where all pilots share the same power as the data symbols. The SBL framework was utilized, based on which a hierarchical Laplace prior was adopted to construct the sparse signal model. The EM algorithm was then employed to update the prior parameters as well as to estimate the channel gains. The work \cite{srivastava2021bayesian} intentionally transmits the pilots in the TF domain, which did not require the guard space in the DD domain, hence increasing the spectral efficiency.
Another work \cite{he2021pilot} commenced from the OTFS modulation and investigated the TF pilot pattern design problem, aiming at minimizing the estimation mean squared error (MSE). The 2D exhaustive pilot location search problem was decoupled into two one-dimensional (1D) problems to reduce the complexity.

Fei et. al proposed a pilot pattern that had a lower proportion of guard symbols \cite{wang2021pilot}. A particle swarm optimization (PSO)-based algorithm was developed for optimizing the OTFS pilot sequence. Finally, interference cancellation scheme is adopted to tackle the pilot contamination issue. Researchers from Southeast University proposed a deterministic pilot design based on the OTFS input-output relationship, which can save both pilot overhead and memory consumption. In \cite{qu2021low}, the authors designed a set of transform domain basis functions to span a low-dimensional subspace for modeling the OTFS channel and proposed a corner-inserted pilot pattern. Then the channel information can be acquired by estimating a few projection coefficients of equivalent channel responses. Recently, two papers \cite{mishra2021otfs} and \cite{yuan2021data} considered a novel pilot-data superimposed frame pattern, which in theory can use the whole frame for data transmission. Iterative channel estimation and data detection protocol was developed to mitigate the interference between pilot and data and to preserve the estimation accuracy.

\subsection{OTFS Detection Design}
In order to achieve the full TF diversity promised by the wireless channel, OTFS modulation generally requires higher detection complexity than the conventional OFDM modulation. This detection complexity increase relates to the separability of the DD domain channel, where the DD domain received signal can be viewed as the superposition of power-attenuated, phase-rotated, and delay- and Doppler-shifted DD domain transmitted signals with respect to each resolvable path's physical attributes of the underlying wireless channel. Consequently, the effective channel matrix for OTFS modulation will have at least $P$ non-zero entries in each row and column, even with integer delay and Doppler shifts~\cite{Raviteja2019practical}. The maximum-likelihood detection corresponding to such a channel matrix will require a detection complexity that grows exponentially with the number of resolvable paths~\cite{li2021hybrid}, which is not desirable in practical applications. Therefore, OTFS detection is a widely studied topic
and is of importance for the practical realization of OTFS systems. 

As discussed in a previous review paper on OTFS detector \cite{zhang2022survey}, the message passing algorithm (MPA) proposed in~\cite{Raviteja2018interference} is commonly adopted for OTFS detection. To achieve a reduced detection complexity, Gaussian approximation is applied to the DD domain intersymbol interference (ISI)~\cite{Raviteja2018LowCom_WCNC}. Furthermore, damping was introduced in the message passing to improve the convergence. Extensive numerical results were provided in~\cite{Raviteja2018interference}, which verify the effectiveness of the proposed MPA.
Similarly, a Gaussian approximation-assisted MPA was introduced in~\cite{Luping2021Gaussian}. In contrast to the MPA in~\cite{Raviteja2018interference}, the Gaussian approximation in this work was applied to the transmitted symbols rather than the interference. Numerical results showed that the proposed MPA can achieve a roughly $1.5$ dB signal-to-noise ratio (SNR) gain compared to the MPA in~\cite{Raviteja2018interference}.
An extension of the MPA, namely, the hybrid maximum \emph{a posteriori} (MAP) and parallel interference cancellation (PIC) detection was proposed in~\cite{li2021hybrid}, where Gaussian approximation was only applied to part of the DD domain ISI depending on the corresponding path attenuation. As a result, this algorithm shows a better error performance compared to the MPA in~\cite{Raviteja2018interference} at the cost of a higher detection complexity. A variational Bayes framework for OTFS detection was introduced in~\cite{Yuan2019simple}. Particularly, an approximation of the joint distribution of OTFS symbols is derived according to the relative entropy. Furthermore, the marginal distribution of a single OTFS symbol is derived based on calculating the variational calculus. Numerical results have demonstrated a better error performance and a superior convergence performance than the MPA in~\cite{Raviteja2018interference}.

Approximated message passing (AMP) is a special type of MPA that generally enjoys a lower complexity. The OTFS detection based on AMP was also studied. A unitary approximate message passing (UAMP)-based OTFS detector was proposed in~\cite{Zhengdao2022UAMP}. This detector is suitable when the number of resolvable paths is large or the fractional Doppler shifts have to be considered. Thanks to the superior performance UAMP, this detector can achieve a promising error performance with an efficient implementation. Furthermore, a UAMP-based OTFS detector operating in the time domain appeared in~\cite{FeiLiu2022MBUAMP}, where the time domain effective channel matrix is partitioned into several small sub-matrices connecting through a factor graph. Thus, the UAMP was applied to handle the estimation within sub-matrices, whose results are iteratively updated according to the factor graph. Numerical results have shown this detector could achieve a promising error performance with practical rectangular pulses.

Other than the MPA-type of detectors, conventional detection methods, such as minimum mean square error (MMSE) detection are also widely applied in OTFS. The low-complexity implementation of MMSE detection for OTFS transmission was investigated in~\cite{Tiwari2019LowComMMSE} and~\cite{Chockalingam2020LowComMMSE}. Both of the studies showed that the MMSE detection can be implemented with a log-linear order of complexity by investigating the properties of the effective channel matrix, such as sparsity and quasi-banded.
A simple two-stage equalizer for OTFS transmission was introduced in~\cite{Chenxi2021simpleTS}. Specifically, a sliding window-assisted MMSE equalizer was first proposed to mitigate the inter-carrier interference (ICI) in the TF domain, and then a simple DD domain equalizer was proposed to eliminate the residual interference from the sliding window-assisted MMSE. Numerical results verified the superior error performance of this two-stage equalizer in comparison to the conventional MMSE detection.
A time domain MMSE and successive interference cancellation (SIC) combined detection was proposed for coded cyclic prefix (CP)-based OTFS in~\cite{Sekhar2021TDCE}. Numerical results have demonstrated that OTFS transmissions are insensitive to synchronization errors under this detection. In the work of \cite{9866655}, the authors introduced interleavers at both transmitter and receiver sides, which gives a sparse matrix form and QR decomposition (QRD) can be readily implemented. Then, the cross-symbol interference was eliminated via SIC. It was shown in the simulations that the QRD-based SIC detectors outperform the non-SIC detectors.

In fact, many existing OTFS detection studies exploited the properties of OTFS effective channel matrices in different domains. These properties offer simple and insightful designs for OTFS detection.
An iterative OTFS detector was proposed under the framework of expectation propagation (EP) algorithm in~\cite{HuaLi2021EP}. In particular, a channel-coefficients-aware scheme was employed during the message passing procedures, which could bundle multiple graph edges into one edge on the factor graph according to the channel coefficients. Numerical results have shown the proposed detector could achieve a good trade-off between performance and complexity. In~\cite{FeiLong2022EP}, an efficient block equalizer for OTFS detection was proposed according to EP. In particular, the authors consider the decomposition of the DD domain effective channel matrix according to the symbol interval, which results in superimposed diagonal-times-doubly circulant matrices. This decomposition reduces the complexity for equalization but does not degrade much about the error performance.
A block-wise OTFS detector was proposed in~\cite{LowCom2021SDIC} by arranging the DD domain transmitted symbols into a block-by-block manner according to the number of delay bins. Based on this symbol arrangement, a least squares minimum residual (LSMR) based channel equalizer was introduced to retrieve the transmitted symbols, which guarantees a quick convergence between the symbol detection and the interference cancellation by leveraging the channel sparsity.
In~\cite{Yao2021TWCFSS}, a fractionally spaced sampling receiver was devised, where the channel input-output relationship with rectangular pulses and oversampling was derived. Based on the derived relationship, two equalization algorithms, namely the iterative combining message passing and the turbo message passing, are introduced. Numerical results have shown that the proposed fractionally spaced sampling receiver can obtain a robust error performance with imperfect channel state information.
In~\cite{Tharaj2020rakeZP}, a low-complexity iterative Rake decision feedback equalizer was proposed for zero-padded OTFS. This equalizer was motivated by the maximal ratio combining (MRC) technology and it can be applied either in DD domain or time domain. A Gauss-Seidel-based over-relaxation parameter was also introduced to improve the convergence and error performances.
The multichannel decision feedback equalizer (DFE) was proposed in~\cite{Lianyou2021CL}, where the spatial diversity was exploited when the channel is experiencing deep fade. Furthermore, an improved proportionate normalized least mean squares algorithm was employed for data detection with fast convergence.
In~\cite{Haojian2021MPDPC}, a low-complexity matched filtering-based message passing detector was proposed, which aims to reduce the high complexity due to the matrix inversion in conventional detection methods. In particular, the probability clipping was adopted in the proposed detector, which can significantly improve the error floor performance at the high SNRs.
A cross domain iterative detection for OTFS was proposed in~\cite{li2021cross}, which is suitable for both the integer and fractional Doppler cases. The key innovation of this detection is that basic estimation/detection approaches are applied to both the time domain and DD domain and the extrinsic information from two domains is iteratively updated according to the unitary transformation. Error performance is analyzed based on the derived state evolution, which shows that this detection can approach the performance of the optimal MAP detection. Inspired by the iterative detection framework, the authors of \cite{haifeng2023tvt} further considered a joint channel estimation and symbol detection scheme based on superimposed pilot-data placement. An orthogonal approximate message passing with the aid of mean-field (MF) message passing was developed for exchanging the soft information between the detector and the channel estimator, bringing enhanced error performance.

The aforementioned studies are focusing on the single-input single-output (SISO) OTFS transmissions. However, the detection for multiple-input multiple-output (MIMO)-OTFS transmissions is more challenging.
A time-space domain channel equalizer was proposed in~\cite{Huiyang2022MIMOOTFS} for MIMO-OTFS detection. Relying on the least squares minimum residual algorithm, this detection algorithm can effectively remove the channel distortion on data symbols. Furthermore, this detection algorithm can be implemented in a recursive manner, where a fast convergence is guaranteed thanks to the sparsity of the MIMO-OTFS channel matrix.
An improved MRC detector was proposed for MIMO-OTFS in~\cite{Tharaj2022MRCMIMOOTFS}, whose main idea was to linearly combine the received signals from different diversity branches (propagation paths and receive antennas). Optimized combing weights were derived. The proposed detector with optimized weights shows a better error performance than the existing detectors for MIMO-OTFS. The error performances of both zero-forcing (ZF) and MMSE were investigated in~\cite{SinghMIMOOTFS}. In particular, the signal-to-interference-plus-noise ratio (SINR) was derived by considering reasonable approximation based on the Taylor expansion, which was then used to provide an accurate estimation on the bit error rate (BER).
A generalized spatial modulation assisted MIMO-OTFS was developed in~\cite{Tiebin2022generalized}, where only part of the transmit antennas are activated to alleviate ICI. A generalized AMP (GAMP) detection algorithm was also introduced to detect the signal, which results in a better error performance than the conventional MMSE detection.

Recent advances in machine learning have also motivated studies for learning-assisted OTFS detection.
A two-dimensional convolutional neural network was introduced in~\cite{Yosef20212D} to facilitate OTFS detection. In particular, the data augmentation technique was considered in this work by leveraging the MPA in~\cite{Raviteja2018interference}. This detection can achieve an improved error performance compared to the MPA, which approaches the optimal MAP detection with low complexity. This work has then been extended to the MIMO case in~\cite{Enku2022ICCW}.
A neural network-based supervised learning framework for OTFS equalization was proposed in~\cite{ZhouZhou2022Learn}. Utilizing reservoir computing, the resulting one-shot online learning is sufficiently flexible to cope with channel variations among different OTFS frames. The main benefit of this framework is that the proposed scheme does not rely on the knowledge of explicit channel state information. Moreover, a better trade-off between the processing complexity and the equalization performance is achieved for OTFS modulation compared to its neural network-based counterparts for OFDM.

\subsection{Multiple Access Designs for OTFS}
The multiple access design is an important subject in practical communication systems. In conventional OFDM transmissions, different users are generally separated by different subcarriers, which is commonly known as the orthogonal frequency division multiple access (OFDMA). However, the success of OFDMA relies deeply on the orthogonality among different subcarriers, which may not hold in the presence of Doppler spread. In light of this, multiple access using OTFS is important, as it can provide a robust performance against the severe Doppler spread.
A unique feature of multiple access in OTFS is the presence of non-negligible multi-user interference (MUI), which is the result of the convolution nature of DD domain channel. Therefore, many recent studies have focused on the multiple access design for OTFS.

A straightforward solution to the DD domain MUI may be the application of guard spaces around each user according to the maximum delay and Doppler shifts. However, this approach generally decreases the data rate due to the increased overhead. In~\cite{Chong2022achievable}, the achievable rates for uplink OTFS-based multiple access without the application of guard space were evaluated and compared with the OFDMA over the high-mobility channels. It was shown in that the OTFS-based multiple access could enjoy a slightly higher achievable rate than OFDMA under the SIC detection, thanks to the channel hardening effect brought by the DD domain communication. As an extension of~\cite{Chong2022achievable}, Chong \emph{et. al} provided a detailed outage analysis for OTFS-based multiple access in~\cite{Ruoxi2022ICC}, which showed that OTFS-based multiple access achieves an improved outage performance than OFDMA.
A special resource allocation scheme was proposed in~\cite{Khammammetti2019OTFSMA}, which ensures the corresponding TF domain signals of different users are not overlapped with each other. Consequently, the MUI is eliminated in the TF domain, resulting in an improved error performance.
In~\cite{Muye2021PDMA}, a new path division multiple access was proposed. By leveraging the beamforming technique, a path scheduling algorithm was introduced to properly assign angle-domain resources for different users such that the observation regions for different users do not overlap over the angle-delay-Doppler domain. Numerical results have demonstrated that the proposed scheme can achieve almost the same error performance as the interference-free case.

Other than the research line focusing on eliminating the MUI, many multiple access designs for OTFS not only embrace the MUI, but also intentionally place the users' data in a non-orthogonal manner to achieve further achievable rate improvement. In~\cite{Zhiguo2019OTFSNOMA}, an OTFS-based non-orthogonal multiple access (NOMA) scheme was introduced, which assigns the user's data to either the DD domain or the TF domain according to their mobility. Sophisticated detection algorithms were also developed to facilitate symbol detection, and their results aligned with the analytical derivations.
An OTFS-based sparse code multiple access (SCMA) was developed in~\cite{Kuntal2021OTFSSCMA}, where a two phases receiver was also proposed for signal detection. Numerical results have demonstrated a better diversity performance over conventional multiple access schemes based on OTFS. Motivated by the fact of principal orthogonality of errors in different domains, \cite{wen2023otfs} extended the cross domain detection to the SCMA system, where OTFS symbol estimation and SCMA decoding were carried out in a joint manner. The cooperation between downlink users was further exploited to develop a distributed cooperative detection algorithm with the aid of belief consensus.

The related research topics for multiple access using OTFS are of importance in practice. Two AMP-based detectors were introduced in~\cite{Geyao2021OTFSNOMA} for signal detection of coded OTFS-NOMA, namely, the orthogonal AMP with linear MMSE (for stationary user) and the Gaussian AMP with EP (for mobile user). Convergence analysis for these two detectors was conducted using the extrinsic information transfer (EXIT) chart, and the advantages of these two detectors over the conventional receiver were verified by numerical results.
A robust beamforming design was introduced to facilitate the application of OTFS-NOMA in~\cite{Zhiguo2020robust}. Particularly, the beamforming problem was formulated to maximize the low-mobility users' data rate while maintaining the targeted data rate of the high-mobility user. Numerical results demonstrated that a significant performance gain can be obtained by the proposed beamforming design in comparison to the random beamforming technology.
The OTFS-NOMA detection using MMSE and SIC was considered in~\cite{Chatterjee2021NOMA}, where low-density parity check (LDPC) codes were applied to facilitate the SIC detection. Both system-level and link-level simulations were developed and compared against the existing multiple access schemes. Numerical results demonstrated that the NOMA-OTFS could provide a higher achievable rate than the orthogonal multiple access schemes based on OTFS. The work of \cite{shi2023screcy} investigateed the security performance of a cooperative multicast-unicast system, where the users exhibit the feature of high mobility and developed the OTFS-NOMA transmission scheme. Furthermore, the authors proposed a power allocation method to strike a trade-off between the reliability of multicast streaming and the confidentiality of unicast streaming and utilized the relay selection strategy to improve the security of unicast streaming. Simulation results showed the superiority of the NOMA-OTFS-based cooperative transmission in terms of both reliability and security for multicast-unicast streaming. The channel estimation for OTFS-SCMA was studied in~\cite{Thomas2022CSC}. Particularly, the channel estimation was developed based on the convolutional sparse coding approach by leveraging the embedded pilot scheme. It was shown that the proposed scheme can maintain a minimal overhead requirement equivalent to a single user without degrading the estimation performance.
A novel random access preamble waveform for OTFS-based uplink multiple access was proposed in~\cite{Sinha2020OTFSMA}, which can efficiently eliminate the round-trip propagation delay between the user and the BS. Numerical results have shown that the proposed preamble waveform enjoys a better timing error probability than the currently deployed counterpart in 4G systems in the presence of Doppler spread.
The user diversity for the OTFS-based random access system was studied in~\cite{Armed2022UD}. The key idea of this work is to assign the available radio resources to the user with the largest number of channel taps, which results in enhanced channel diversity. Extensive numerical results have demonstrated that the proposed scheme outperforms the conventional OTFS-based random access in terms of the BER.

\section{OTFS-enabled Integrated Sensing and Communication}
Environment sensing, in addition to efficient communications, is also highly demanded in the 6G era. Although for simplicity, researches on radar sensing and communication were on two parallel streams and separate frequency bands were allocated for different functionalities, the underutilization of the system resources is not sustainable for the development of long-term wireless networks. To this end, there is a recent trend to develop the integrated sensing and communication (ISAC) system, which allows the reuse of the same frequency resources, signal processing framework, and hardware architectures for both functionalities \cite{liu2022integrated,cui2021integrating}. With the aid of ISAC technology, the resource efficiency as well as the system throughput can be improved \cite{chiriyath2017radar,liu2020joint}.

To realize ISAC, pioneering works considered the adoption of OFDM signaling, which offers satisfactory performance for both sensing and communications. Note that the sensing-related applications inherently focus on the delay and Doppler parameters representing the range and velocity characteristics of moving targets, which perfectly aligns with OTFS. In contrast to OFDM-based ISAC systems, OTFS-ISAC systems provide a direct relation between the transmitted signals and the channel response in one unified DD domain for both functionalities, in addition to their resilience to delay and Doppler spreads. For this reason, OTFS has been regarded as a promising signal waveform for fully unleashing the potential of ISAC systems. The work \cite{raviteja2019orthogonal} compared the radar sensing performance of OTFS and OFDM waveforms, showing that OTFS achieves a more accurate velocity estimation. G. Caire et.al were the first in the literature to analyze the OTFS-based sensing and communication performances \cite{gaudio2020effectiveness}. Maximum likelihood estimator and Cramér-Rao lower bound for range and velocity estimation were derived. The results showed that for the dual-functional sensing and communications use cases, OTFS signaling can provide as accurate parameter estimation as the dedicated radar waveforms, e.g., FMCW. Moreover, OTFS performs better than OFDM in terms of channel capacity and overhead cost. The work of \cite{keskin2021radar} derived a new OTFS radar signal model by taking into account the ISI and ICI. Benefiting from the new signal model, both ICI and ISI phenomena can be turned into an advantage to surpass the maximum radar unambiguous detection limits. The authors of \cite{yuan2021integrated} presented a framework of OTFS-based ISAC systems in vehicular networks. By adopting the slow time-variation of OTFS channel, the delay and Doppler shifts obtained from the sensing echoes can be used for inferring the communications channel, which is also characterized by delays and Dopplers. Consequently, a new data frame structure was designed which can make full use of the delay/Doppler grids and the overhead for channel estimation was reduced. An iterative receiver was proposed to recursively update the channel information and detect data symbols. K. Wu et. al considered OTFS-based ISAC systems in industrial Internet-of-Things (IIoT) \cite{wu2021otfs}. Using a series of waveform pre-processing, the impact of communication data symbols was removed in the TF domain. A high-accuracy and off-grid method was then proposed for estimating range and velocity, which experiences a very low complexity dominated by two-dimensional DFT. Further SINR analysis was derived for the proposed algorithm. Extensive simulations validated the near-optimal sensing performance and high energy efficiency of OTFS in IIoT applications. A recent work on this topic by Li et. al developed a spatial spread OTFS (SS-OTFS)-based ISAC framework \cite{li2022novel}. With spatial spreading/de-spreading enabled discretized angular domain, insightful effective models for communications and radar sensing can be derived. In particular, three use cases, i.e., beam tracking, power allocation, and angle estimation are studied with a special structure of sensing matrix. The most important conclusion in this paper is that the power allocation should be designed leaning toward sensing in a practical SS-OTFS-based ISAC system.

Relying on the same DD domain for both communications and sensing, OTFS offers additional promising benefits for ISAC systems, which has been discussed in \cite{yuan2022otfstutorial} and is summarized as follows,
\begin{itemize}
	\item As both communications and sensing signals are represented in the DD domain, one can write the acquired sensing echo and the received communications OTFS signal in concise form as $\mathbf{y} = \mathbf{H} \mathbf{x} +\mathbf{n}$, where $\mathbf{H}$ and $\mathbf{x}$ denote the channel and transmitted signal, respectively. For sensing, the transmitted signal is known and the goal is to infer sensing channel. While for communications, the data symbols are to be detected given the CSI. The sensing performance is characterized by the ambiguity function and the CRB while the communication performance is evaluated in terms of the channel capacity and BER. Obviously, the optimal signal design for communication and sensing may not be the same. The OTFS waveform enables us to formulate the signal design problem in the same DD domain. For example, in a communication-centric scenario, we aim to maximize the channel capacity, while maintaining a desired sensing performance.
\item The OTFS channel model reflects the underlying geometry of the wireless propagation environment. After channel estimation, the estimated delays and Doppler shifts can be converted to range distances and speeds. Given a temporal sequence of estimated delays corresponding to the $i$-th path, the location of the $i$-th reflector can be obtained via an estimator.
If the reflector is also a moving target, Bayesian estimators such as extended Kalman filtering (EKF) can be used to exploit the state transition information. By combining the sensing information gleaned from the reflected echoes at the BS and OTFS channel estimation, we can construct a more precise image of the environment.
\end{itemize}

\section{Emerging OTFS Applications}
Due to the benefits of low-complexity implementation, sparse channel representation, and high resilience to large delays and Doppler spreads, OTFS is attractive for several emerging applications.

Aiming for huge and unregulated bandwidth in future wireless systems, the optical spectrum is of significant interest since it is not affected by existing electromagnetic interference. For visible light communication (VLC), the data symbols are transmitted using light-emitting diodes (LED) by adapting the luminous intensity. At the receiver side, the photodiode acquires the signal and converts it to an electrical one. VLC channel is generally assumed to be static and multi-path propagation. As discussed in Sec III, OTFS is also a good choice for static multi-path channels due to its capability of exploiting full TF diversity. The paper \cite{sharma2020performance} considered a single LED system and showed that OTFS-based VLC can outperform its OFDM-based counterpart. The work of \cite{zheng2021dco} proposed a direct current biased optical OTFS (DCO-OTFS) based full-duplex relay-assisted VLC system. With the benefits of two-dimensional domain multiplexing, OTFS can reduce the number of cyclic prefix than OFDM modulation, which enhances the spectral efficiency of VLC. This is the first work applying OTFS into a VLC system with frequency domain relaying, which provides guidelines for the system design of an OTFS-based VLC system. The work of \cite{zhong2020orthogonal} proposed a new orthogonal time-frequency multiplexing (OTFM) scheme for DC-biased VLC based on OTFS modulation and designed a 2D Hermitian symmetry in the DC-biased optical system to generate the real-valued signal. Numerical results demonstrated its superiority over the OFDM scheme.

In underwater environments, acoustic waves instead of radio signals are usually utilized for communications purposes. The low signal propagation speed of acoustic waves results in severe delay and Doppler spreads for underwater acoustic (UWA) communications. The delay could spread up to hundreds of milliseconds while the Doppler shift is also very large. Moreover, the multi-path environment is complex due to numerous reflectors from the sea surface and sea floor. In contrast to the OFDM scheme, OTFS technique can benefit from the DD channel representation and enable efficient channel estimation/detection. The paper \cite{feng2020underwater} first time considered the application of OTFS in UWA communications and provide the detailed OTFS-based UWA communications model. Simulation results show that the OTFS-based scheme has shown significant performance gain compared to OFDM system under time-varying UWA channels. Jing et.al proposed a two-dimensional adaptive multi-channel decision feedback equalization (2D DFE) to utilize the spatial gain and achieve good performance \cite{jing2022otfs}. The authors of \cite{zhang2023channel} considered OTFS-UWA system and formulated a sparse channel estimation problem for the generalized linear models (GLM). A structured sparsity-based GAMP algorithm was proposed for reliable channel estimation in quantized OTFS systems to leverage the structured sparsity of the doubly spread UWA channel. Experimental results were also conducted to show accurate UWA channel estimation result. 

Index modulation (IM) technique, which considers innovative ways to convey information has emerged as a promising technology
for next-generation wireless networks due to the advantages of energy efficiency and hardware simplicity \cite{basar2017index}. A new transmission scheme, namely OTFS with dual-mode index modulation (OTFS-DM-IM) was proposed in \cite{zhao2021orthogonal} to balance the reliability and spectral efficiency. The BER performance of the proposed transmission scheme was analyzed, which showed the performance advantage over classic OTFS systems. In \cite{feng2022phase}, OTFS with in-phase and quadrature index modulation was proposed to further increase the spectral efficiency. In particular, the additional information bits were embedded in the grid index along in-phase and quadrature dimensions. The maximum PAPR was derived to evaluate the performance. The work of \cite{li2023iterative} presented an iterative receiver for coded OTFS-IM system. The authors devised the structured prior incorporating activation pattern constraint and channel coding and develop an iterative receiver via structured prior-based hybrid belief propagation (BP) and EP algorithm. Then, two variations of the proposed algorithm were derived using some approximations.

As for the realization of SAGIN, there has not been a single paper discussing OTFS-based SAGIN specifically. However, several contributions have considered the applications of OTFS in LEO, UAV, and vehicular networks. Given the bottleneck of supporting large areas and high mobility speeds using the current technological regime, \cite{xu2020research} studied OTFS modulation under the parameters of LTE-based terrestrial broadcast. Simulation results regarding OTFS system under the numerologies of terrestrial broadcast published in R14 were also provided. In \cite{wang2022joint}, the authors proposed a joint channel estimation and data detection method for OTFS-based LEO satellite communications. The 'to-be-detected' data symbols served as the virtual pilots, which helps to improve the channel estimation accuracy. In \cite{hu2021secrecy}, the authors derived the expression of security outage probability (SOP) at the legitimate receiver in a closed-form, where the summed pdf of the shadowed-Rician distributed terms was found through moment matching. The work of \cite{shi2022outage} considered UAV-based cooperative transmission for compensating the large path-loss caused by long-distance satellite communications. The conclusion was drawn that OTFS significantly outperforms the OFDM scheme, which can strike a trade-off between reliability improvement and implementation complexity determined by modulation size. Han et al. investigated an OTFS-based UAV communication system in \cite{han2023trajectory} to tackle the challenges introduced by large path-loss and severe Doppler effect due to the high mobility of UAVs. Through simulations, OTFS is shown to accomplish transmission tasks over shorter distances with lower energy consumption compared to OFDM. In \cite{linsalata2021otfs}, the authors proposed a localization method based on UAVs. In particular, a fully standards-compliant OTFS-modulated physical random access channel {PRACH} transmission and reception scheme to perform time-of-arrival (ToA) measurements. In high-speed railway communications, the environment varies frequently, imposing challenges for reliable communications. In \cite{ma2021otfs}, the authors jointly designed tandem spreading multiple access (TSMA) and OTFS for high-speed railway communications. The principle of OTFS-TSMA transceiver was presented, where OTFS and TSMA were improved respectively. The authors of \cite{shang2023otfs} investigated the reliable communication in IoT for railways (IoT-R) under high mobility scenarios and strict energy constraints. In particular, the peak windowing technique was used to decrease the PAPR of OTFS and promote the application of OTFS modulation in IoT-R. For cellular systems on millimeter-wave and terahertz spectrum, the authors from Samsung proposed compressed sensing-based OTFS transmission, where a multi-dimensional sparse vector compression scheme converts a non-sparse transmit vector to a sparse one \cite{ji2021compressed}. Then, DD domain spreading was applied by using only a few TF grids. The proposed algorithm outperforms the existing works using only $10$ - $40$\% of frequency resources compared to conventional schemes.

\section{Research Challenges and Future Works}
In the above, we have reviewed the literature of OTFS, including the basic concepts, fundamental performance, transmitter, and receiver design, DD signal processing-enabled ISAC, and emerging applications. However, as a new waveform technology, several aspects of OTFS modulation have not been well addressed by the research community yet. Here, we summarize some open problems as follows:

\subsection{Low Latency OTFS Transmission}
Despite the advantages of combating channel fluctuations, OTFS may experience higher detection latency than OFDM due to block-based data detection. On the other hand, the quasi-static and sparse DD channel can significantly reduce overhead cost and computational complexity. Therefore, the overall latency of OTFS and OFDM still requires further study. The performance comparison of OTFS and OFDM in terms of communication latency is an interesting topic. Moreover, although OTFS utilizes a full block for data transmission, the received samples in the time domain actually contain all transmitted information bits. To this end, AI-based detection frameworks may exploit the correlation between consecutive received samples and detect OTFS symbols with lower latency.

\subsection{Insufficient Doppler/delay resolutions}
To partially address the latency issue discussed in the previous subsection, we can limit the signal duration, which, however, imposes a fractional Doppler effect due to insufficient Doppler resolution. As for the delay, fractional delay also appears if the bandwidth is narrow. Various approaches have been proposed to address the DD channel estimation in the presence of fractional delay/Doppler, such as adding a window function and off-grid estimation. However, reliably detecting the data symbols in these use cases is a critical issue. The commonly used MPA detector may suffer from performance degradation because of the dense factor graph. New detection schemes that are robust to fractional Doppler/delay are of great interest.

\subsection{Multi-user and multi-antenna OTFS}
Since OTFS was first proposed, serving multiple users has always been a topic of great interest. Several research works, as reviewed in Section III of this article, have attempted to address this issue. However, it remains unclear which domain users should be multiplexed in. The 2D circular convolution of transmitted symbols and effective channels in the DD domain results in a unique inter-user interference (IUI) pattern. Thus, we may design DD domain user multiplexing schemes from the perspective of minimizing IUI. Furthermore, multiple-input multiple-output (MIMO) systems can be exploited to achieve higher achievable rates. Nevertheless, joint signaling optimization in delay, Doppler, and spatial domains is highly complex. Considering the multi-user case, a general and concise multi-user MIMO-OTFS model is anticipated and under research.

\subsection{Cross Layer Design}
Future high-mobility wireless applications will not only involve rapid time fluctuations but also dynamic network topologies. As a result, cross-layer joint design, which encompasses both physical and network layers, presents an intriguing yet challenging task. Network topology information may aid OTFS waveform and transceiver design. Moreover, to fulfill specific communication requirements, it might be possible to further optimize network topology based on the wireless environment.

\subsection{Delay Doppler Channel Exploitation for ISAC}
As discussed in \cite{yuan2021integrated}, the delay and Doppler parameters associated with the sensing channel were reused for OTFS communication channel prediction, where all targets of interest are identical to the communication users. However, this is not always the case. Depending on the scenario, only part of the targets in the area of coverage may be UEs. Therefore,  fundamental question arises: how much information can we glean from the sensing channel to infer the communications channel? The first step involves characterizing the similarity between the OTFS communication and sensing channels. A natural choice is to use the cross-correlation matrix, where its entries reflect the correlation between sensing targets and communication reflectors corresponding to different paths. Although OTFS modulation offers the opportunity to connect both channels in the same DD domain, new analysis tools and frameworks still need to be developed to further study the exploitation of the sensing channel in OTFS-ISAC systems.

\subsection{OTFS-aided RIS/Backscatter Communications}
RIS/backscatter communications have attracted significant attention for simultaneously supporting active primary transmission and potential passive information transmission. Specifically, their key idea is to modify and reflect original signals to save on hardware costs and energy consumption. OFDM-aided RIS/backscatter communications have been intensively investigated in the literature but are not suitable for future time-varying wireless scenarios. Naturally, researching OTFS-aided RIS/backscatter communications holds great potential. Some pioneering researches have been conducted recently \cite{bhat2022input,zhongjie2023iot}, where input-output relationships, data detection, and channel estimation were discussed. Nevertheless, the benefits of OTFS-RIS communications have not been thoroughly studied yet.

\subsection{Predictive Delay Doppler Communications}
The slow time-varying nature of the DD domain channel offers valuable insights into predictive DD communications. Given communication transceivers moving at approximately constant speeds, the delay and Doppler shifts associated with the channel remain nearly constant. Consequently, the channel can be estimated only once to facilitate the transmission of multiple OTFS frames. Another potential research direction is the development of DD predictive precoding design. An initial attempt has been made in \cite{liu2023predictive}, where a deep learning approach was proposed to exploit implicit features from estimated historical DD channels, enabling the direct prediction of the precoder to be employed for minimizing the error rate in the subsequent frame transmission. Relying on the powerful deep learning framework, the ultimate goal is to completely bypass channel estimation for DD communications, which represents a highly promising topic for future research endeavors.

\section{Conclusions}
Delay-Doppler communications relying on OTFS has been seen as a promising enabler for future wireless systems, particularly in high-mobility environments. This article presents a comprehensive survey of OTFS technology in the context of the 6G era. Specifically, we begin with the background of DD domain channel representation, followed by an examination of the fundamentals and performance limits associated with OTFS. As the most critical aspect, we provide a detailed account of the latest research advancements in transmitter and receiver design. Subsequently, we discuss the OTFS-based ISAC system. Lastly, we summarize several emerging applications of OTFS and identify open research challenges. Our aim is for this survey paper to assist researchers in both industry and academia to gain a deeper understanding of OTFS and to encourage further exploration in this area.
\label{Research}

\section*{Acknowledgements}
This work is supported in part by National Natural Science Foundation of China under Grant 62101232, in part by Guangdong Provincial Natural Science Foundation under Grant 2022A1515011257, and in part by Shenzhen Science and Technology Program under Grant JCYJ20220530114412029.

\bibliographystyle{gbt7714-numerical}
\bibliography{myref}

\end{document}